\documentclass[12pt]{article}
\usepackage{graphicx}
\def \b{{\cal B}}

\def \beq{\begin{equation}}
\def \beqn{\begin{eqnarray}}
\def \bo{B^0}
\def \eeq{\end{equation}}
\def \eeqn{\end{eqnarray}}
\def \cn{Collaboration}
\def \ite{{\it et al.}}
\def \ob{\overline{B}^0}

\def \ok{\overline{K}^0}
\def \s{\sqrt{2}}

\def \st{\sqrt{3}}
\def \sx{\sqrt{6}}
\def \tG{\tilde{\Gamma}}
\def \v#1#2{V_{#1#2}}

\begin{document}
\renewcommand{\thetable}{\Roman{table}}
\rightline{ANL-HEP-PR-02-113}
\rightline{EFI-02-46-Rev}
\rightline{hep-ph/0212274}
\rightline{December 2002}
\bigskip
\bigskip
\centerline{\bf FINAL-STATE PHASES IN $B \to D \pi,~D^* \pi$, AND}
\centerline{\bf $D \rho$ DECAYS
\footnote{To be submitted to Phys.~Rev.~D.}}
\bigskip
\centerline{\it Cheng-Wei Chiang}
\centerline{\it High Energy Physics Division, Argonne National Laboratory,
Argonne, IL 60440}
\centerline{and}
\centerline{\it Enrico Fermi Institute,
University of Chicago, Chicago, IL 60637}
\medskip

\centerline{\it Jonathan L. Rosner}
\centerline{\it Enrico Fermi Institute and Department of Physics}
\centerline{\it University of Chicago, Chicago, IL 60637}
\bigskip
\centerline{\bf ABSTRACT}
\medskip
\begin{quote}
  The final-state phases in $\overline{B} \to D \pi,~D^* \pi$, and $D \rho$
  decays appear to follow a pattern similar to those in $D \to \overline{K}
  \pi$, $\overline{K}^* \pi$, and $\overline{K} \rho$ decays.  Each set of
  processes is characterized by three charge states but only two independent
  amplitudes, so the amplitudes form triangles in the complex plane.  For the
  first two sets the triangles appear to have non-zero area, while for the $D
  \rho$ or $\overline{K} \rho$ decays the areas of the triangles are consistent
  with zero.  Following an earlier discussion of this behavior for $D$ decays,
  a similar analysis is performed for $B$ decays, and the relative phases and
  magnitudes of contributing amplitudes are determined.  The significance of
  recent results on $\ob \to D^{(*)0} \overline{K}^{(*)0}$ is noted.  Open
  theoretical and experimental questions are indicated.
\end{quote}
\medskip
\leftline{\qquad PACS codes: 13.25.Hw, 11.30.Hv, 14.40.Nd, 13.75.Lb}
\bigskip

\centerline{\bf I. INTRODUCTION}
\bigskip

The decays of $B$ mesons (those containing the $b$ quark) are potentially
rich sources of information on CP violation.  One such manifestation of this
phenomenon involves an asymmetry $A(f)$ between the rate for a decay of a
$B$ meson to a final state $f$ and the corresponding CP-conjugate
process:
\beq
A(f) \equiv \frac{\Gamma(B \to f) - \Gamma(\bar B \to \bar f)}
{\Gamma(B \to f) + \Gamma(\bar B \to \bar f)}~~~.
\eeq
Such an asymmetry requires there be at least two contributing amplitudes
$A_{1,2}$, each characterized by distinct weak phases $\phi_{1,2}$
and strong phases $\delta_{1,2}$.  Under CP-conjugation, the weak phases
change sign but the strong phases do not:
\beq
A(B \to f) = |A_1|e^{i \phi_1} e^{i \delta_1} +
             |A_2|e^{i \phi_2} e^{i \delta_2}~~~,
\eeq
\beq
A(\bar B \to \bar f) = |A_1|e^{-i \phi_1} e^{i \delta_1} +
             |A_2|e^{-i \phi_2} e^{i \delta_2}~~~,
\eeq
so that $A(f) \propto \sin(\phi_1-\phi_2) \sin(\delta_1 - \delta_2)$.  In
these decays the observation of a so-called ``direct'' CP asymmetry thus
requires both the weak and the strong phases of the two contributing
amplitudes to differ from one another.  Thus it is of great importance to
understand the patterns of strong final-state phases in as wide as possible a
set of decays.

The strong final-state phases in decays of strange particles are appreciable.
For example, in $K_{S,L} \to \pi \pi$ the final-state phases in the $I_{\pi
  \pi} = 0$ and $I_{\pi \pi} = 2$ channels differ from one another by many tens
of degrees.  Furthermore, they can be measured directly in elastic $\pi \pi$
scattering, and then applied to the decays $K_{S,L} \to \pi \pi$ using Watson's
Theorem \cite{Watson}.  However, in the decays of charmed and heavier mesons to
two-body final states, these states constitute only a small fraction of the
available decays, and elastic phase shifts are no longer relevant
\cite{WolfFSI,DGPS}.

In the limit of a very heavy decaying quark, certain nonleptonic decays are
expected to be characterized by small final-state interactions.  These are the
ones such as $\bar B^0 \to D^{(*)+} \pi^-$ to which the {\it factorization}
hypothesis \cite{BSW,BJ} applies: The decay amplitude can be regarded as the
product of two color-singlet currents, one associated with the $\bar B^0 \to
D^{(*)+}$ transition and the other creating the $\pi^-$ from the vacuum.  The
large relative momentum of the two final-state particles leaves little time for
them to interact with one another before they are safely out of each other's
range.  This expectation is confirmed in recent analyses of the factorization
hypothesis based on QCD \cite{BBNS}, though the role of final-state
interactions in other nonleptonic heavy quark decays is more open to question
(see, e.g., \cite{KLS}).

Some processes involve {\it color-suppressed} weak decays, in which the weak
current produces a pair of quarks each of which ends up in a different meson.
Other processes involve interactions in which the quark and antiquark in the
initial meson annihilate with one another or exchange a $W$ boson.  While these
last processes are expected to be suppressed by a factor of (decay
constant)/(heavy meson mass) in the amplitude relative to those to which
factorization should apply, they have been found not to exhibit such
suppression in charmed meson decays \cite{JRcharm}.  As a result, charmed
particle decays exhibit an interesting pattern of final-state phases, in which
there are large relative phases between the amplitudes for the various charge
states in $D \to \overline{K}\pi$ and $D \to \bar K^* \pi$, but the amplitudes
for $D \to \overline{K}\rho$ seem to be relatively real with respect to one
another \cite{MkIII,Kamal,Banff,SuzDB,JRFSI,Anjos,ARGUS,Frab}.

The decays $\bar B \to D \pi,~D^* \pi, D \rho$ now have been studied with
sufficient accuracy that a similar pattern appears to be emerging.  The rates
for the three charge states in the first two processes favor relative phases
between contributing amplitudes (e.g., those of definite isospin)
\cite{PedlarD,Ahmed}, while the $\bar B \to D \rho$ rates favor amplitudes
which are relatively real \cite{BelDrho}.  In the present paper we perform an
analysis parallel to that for charmed particles in Ref.\ \cite{JRcharm},
finding that the source of the final-state phases in the $\bar B$ decays under
discussion is very similar to that in charm decays, but that the effects are
diminishing as expected with increasing heavy quark mass.  We point out open
theoretical and experimental problems and indicate what further data would be
useful in resolving them.

Now that the color-suppressed decays $\ob \to D^0 \ok$ and $\ob \to D^0
\overline{K}^{*0}$ have been observed \cite{BeDK}, one can perform a
similar analysis for $\overline{B} \to D \overline{K}$ decays.  However,
in contrast to a recent claim \cite{Xing}, we find that the experimental
errors on these Cabibbo-suppressed decay modes are still too large to permit
any firm conclusion about relative strong phases.  We shall discuss the
importance of such modes in reducing ambiguities in the Cabibbo-favored
amplitudes.

We review the experimental situation for $\bar B \to D \pi,~D^* \pi, D \rho$ in
Sec.\ II, performing a standard isospin analysis and confirming that the
isospin amplitudes have a non-zero relative phase for the first two processes
but not for the third.  We then introduce a description of the decays in terms
of topological amplitudes in Sec.\ III.  The implications of the data for these
amplitudes are discussed in Sec.\ IV, while we discuss missing pieces of the
puzzle and experimental prospects in Sec.\ V.  We summarize in Sec.\ VI.

\bigskip
\centerline{\bf II.  EVIDENCE FOR RELATIVE PHASES IN (SOME) $\bar B$ DECAYS}
\bigskip

We review the isospin decomposition for the decays $\overline{B} \to D \pi$,
following closely the corresponding discussion for $D \to \overline{K} \pi$
\cite{JRFSI}.  Similar decompositions then follow when one of the final-state
particles is a vector meson, since in all cases there is a single partial wave
in the decay.

The decays of interest are governed by the subprocess $b \to c d \bar u$, which
has $\Delta I = 1$, $\Delta I_3 = -1$.  Since the initial $\overline{B}$ state
has $I = 1/2$, the processes are characterized by two amplitudes $A_{1/2}$ and
$A_{3/2}$ labeled by the total isospin of the final $D \pi$ state.  The
amplitudes are given by
$$
{\cal A}(D^0 \pi^-) = A_{3/2}~~,
$$
$$
{\cal A}(D^+ \pi^-) = \frac{2}{3} A_{1/2} + \frac{1}{3} A_{3/2}~~,
$$
\beq
\label{eq:amps}
{\cal A}(D^0 \pi^0) = - \frac{\s}{3}A_{1/2} + \frac{\s}{3}A_{3/2}~~,
\eeq
where we omit the initial particle.  They thus satisfy a triangle relation
\beq
{\cal A}(D^0 \pi^-) = {\cal A}(D^+ \pi^-) + \s {\cal A}(D^0 \pi^0)~~.
\eeq
A non-zero area of the triangle would signify non-trivial final-state
phases between the two isospin amplitudes. 

Letting $\Phi_{i}$ denote kinematic factors which we shall specify shortly,
where the subscript denotes the final state,
we can define reduced partial widths with the kinematic factors removed, e.g.,
\beq
|A_{3/2}|^2 = \tG(D^0 \pi^-) \equiv \Gamma(D^0 \pi^-)/\Phi_{D^0 \pi^-}~~,
\eeq
\beq
|A_{1/2}|^2 = \frac{3}{2} [ \tG(D^+ \pi^-) + \tG(D^0 \pi^0) ]
- \frac{1}{2} \tG(D^0 \pi^-)~~,
\eeq
and the relative phase $\delta_I = {\rm Arg}(A_{3/2}/A_{1/2})$ between isospin
amplitudes satisfies
\beq
\label{eq:cos}
\cos \delta_I = \frac{3 \tG(D^+ \pi^-) + \tG(D^0 \pi^-) - 6 \tG(D^0 \pi^0)}
{4 |A_{1/2} A_{3/2}|}~~~.
\eeq

The search for relative phases in $\overline{B} \to D \pi,~D^* \pi$, and $D
\rho$ decays goes back at least as far as the unpublished work of Yamamoto
\cite{HY}, in which only upper limits existed at the time for the
color-suppressed decays $\bo \to D^0 \pi^0,~D^{*0} \pi^0$ and $D^0 \rho^0$.
Belle and CLEO reported observation of the first two final states about a year
and a half ago.  The rate for $\ob \to D^0 \pi^0$ was found to be large enough
that the triangle of complex amplitudes for $B^0 \to \bar D^0 \pi^0$, $B^0 \to
D^- \pi^+$, and $B^+ \to \bar D^0 \pi^+$ appeared to have non-zero area.  More
recently, CLEO reported an analysis of a larger data sample of the last two
modes \cite{PedlarD,Ahmed}, which strengthens the argument for a non-zero final
state phase difference between the $I = 1/2$ and $I = 3/2$ amplitudes.  With
the new branching ratios $\b(\ob \to D^+ \pi^-) = (26.8 \pm 2.9) \times
10^{-4}$, $\b(B^- \to D^0 \pi^-) = (49.7 \pm 3.8) \times 10^{-4}$ as well as
the Belle-CLEO average $\b(\ob \to D^0 \pi^0) = (2.92 \pm 0.45) \times
10^{-4}$, one finds $|A_{3/2}|=(7.70 \pm 0.29) \times 10^{-7}$ GeV, $|A_{1/2}|
= (5.30 \pm 0.58) \times 10^{-7}$ GeV, and $\cos \delta_I = 0.86 \pm 0.05$ or
$\cos \delta_I < 1$ at $2.8 \sigma$.

The same formulae (\ref{eq:amps})--(\ref{eq:cos}) can be readily applied
to both $D^* \pi$ and $D \rho$ decays by replacing the final state mesons with
the appropriate ones.  It is found that the $D^* \pi$ decays have $|A_{3/2}| =
(3.32 \pm 0.14) \times 10^{-7}$, $|A_{1/2}| = (2.50 \pm 0.20) \times 10^{-7}$,
and $\cos \delta_I = 0.86 \pm 0.06$ or $\cos \delta_I < 1$ at $2.4 \sigma$.
Similar conclusions were drawn for isospin amplitudes in $\overline{B} \to
D^{(*)} \pi$ in Ref.\ \cite{NP}.

$D^* \pi$ decays thus have a phase structure very similar to that of the $D
\pi$ decays.  However, with the newly reported branching ratio ${\cal B}(\ob
\to D^0 \rho^0) = (2.9 \pm 1.0 \pm 0.4) \times 10^{-4}$ \cite{BelDrho}, the $D
\rho$ decays give $|A_{3/2}| = (5.74 \pm 0.39) \times 10^{-7}$, $|A_{1/2}| =
(4.00 \pm 0.76) \times 10^{-7}$, and $\cos \delta_I = 0.99 \pm 0.08$,
consistent with a vanishing strong phase.

The decays $\overline{B} \to D^{(*)} \overline{K}^{(*)}$ are governed by
the quark subprocess $b \to c \bar u s$, which has $\Delta I = - \Delta I_3
= 1/2$.  Combining this interaction with the isospin $I=1/2$ of the initial
$\overline{B}$, one has two amplitudes $A^{DK}_0$ and $A^{DK}_1$ labeled by
total isospin.  For example, for $\overline{B} \to D \overline{K}$, using the
phase convention of Ref.\ \cite{Xing},
$$
{\cal A}(\ob \to D^+ K^-) = \frac{1}{2}A^{DK}_1 + \frac{1}{2}A^{DK}_0~~,~~~
{\cal A}(\ob \to D^0 \ok) = \frac{1}{2}A^{DK}_1 - \frac{1}{2}A^{DK}_0~~,~~~
$$
\beq
{\cal A}(B^- \to D^0 K^-) = A^{DK}_1~~~.
\eeq
One thus has the sum rule
\beq \label{eqn:sr}
{\cal A}(\ob \to D^+ K^-) + {\cal A}(\ob \to D^0 \ok) =
{\cal A}(B^- \to D^0 K^-)~~,
\eeq
with similar sum rules when one final pseudoscalar is replaced by a vector
meson.  (When both final mesons have spin 1, there are three helicity
amplitudes or partial waves; the sum rule holds for each.
We shall not consider such decays further here.)

Xing \cite{Xing} has argued that the observed amplitudes for $D \overline{K}$
and $D \overline{K}^*$ decays favor non-zero relative phases between the
isospin amplitudes.  We shall see that these amplitudes are consistent with
being relatively real at better than $1 \sigma$, and will identify the
improvements in measurements that are likely to be needed in order to
establish a non-zero relative phase.
\bigskip

\centerline{\bf III.  TOPOLOGICAL AMPLITUDES}
\bigskip

Meson wave functions are assumed to have the following quark content, with
phases chosen so that isospin multiplets contain no relative signs
\cite{GHLR,eta}:

\begin{itemize}

\item{\it Beauty mesons:} $\ob = b \bar d$, $B^- = - b \bar u$,
$\overline{B}_s = b \bar s$.
 
\item {\it Charmed mesons:} $D^0 = - c \bar u$, $D^+ = c \bar d$, $D_s^+ =c
  \bar s$, with corresponding phases for vector mesons.
  
\item {\it Pseudoscalar mesons $P$:} $\pi^+ = u \bar d$, $\pi^0 = (d \bar d - u
  \bar u)/\s$, $\pi^- = - d \bar u$, $K^+ = u \bar s$, $K^0 = d \bar s$, $\bar
  K^0 = s \bar d$, $K^- = - s \bar u$, $\eta = (s \bar s - u \bar u - d \bar
  d)/\st$, $\eta' = (u \bar u + d \bar d + 2 s \bar s)/\sx$, assuming a
  specific octet-singlet mixing \cite{Chau,eta} in the $\eta$ and $\eta'$ wave
  functions.)
  
\item {\it Vector mesons $V$:} $\rho^+ = u \bar d$, $\rho^0 = (d \bar d - u
  \bar u)/\s$, $\rho^- = - d \bar u$, $\omega = (u \bar u + d \bar d)/\s$,
  $K^{*+} = u \bar s$, $K^{*0} = d \bar s$, $\overline{K}^{*0} = s \bar d$,
  $K^{*-} = - s \bar u$, $\phi = s \bar s$.

\end{itemize}

The partial width $\Gamma$ for a specific two-body decay to $PP$ is expressed
in terms of an invariant amplitude ${\cal A}$ as 
\beq
\Gamma(B \to PP) = \frac{p^*}{8 \pi M^2}|{\cal A}|^2~~~,
\eeq
where $p^*$ is the center-of-mass (c.m.) 3-momentum of each final particle,
and $M$ is the mass of the decaying particle.  The kinematic factor of $p^*$ is
appropriate for the S-wave final state.  The amplitude ${\cal A}$ will thus
have dimensions of (energy)$^{-1}$.

For $PV$ decays a P-wave kinematic factor is appropriate instead, and
\beq
\Gamma(B \to PV) = \frac{(p^*)^3}{8 \pi M^2}|{\cal A'}|^2~~~.
\eeq
Here ${\cal A'}$ is dimensionless.  These conventions agree with those of Chau
\ite~\cite{Chau}.

The amplitudes ${\cal A}$ are then expressed in terms of topological amplitudes
of three types.

\begin{itemize}
  
\item {\it Tree amplitudes $T$:} These are associated with the transition $b
  \to c d \bar u$ (favored) or $b \to c s \bar u$ (suppressed) in which the
  light (color-singlet) quark-antiquark pair is incorporated into one meson,
  while the charmed quark combines with the spectator antiquark to form the
  other meson.  We denote (favored, suppressed) amplitudes by (unprimed,
  primed) quantities, respectively.

\item {\it Color-suppressed amplitudes $C$:} The transition is the same as in
  the tree amplitudes, namely $b \to c d \bar u$ or $b \to c s \bar u$, while
  the charmed quark and the $\bar u$ combine into one meson while the light
  quark and the spectator antiquark combine into the other meson.
  
\item{\it Exchange amplitudes $E$:} The $b$ and spectator antiquark exchange a
  $W$ to become a $c \bar u$ pair, which then hadronizes through the creation
  of a light quark-antiquark pair.

\end{itemize}

We neglect a fourth type of (annihilation) transition in which a $b$ and a
$\bar u$ annhiliate to form an $s \bar c$ or $d \bar c$ pair.  Such transitions
do not contribute in any case to $\bar B \to D + X$ decays.

For reference, the relation between isospin amplitudes and topological ones
for Cabibbo-favored decays is
\beq
A_{3/2} = {\cal A}(D^0 \pi^-) = - (T+C)~~,~~~
A_{1/2} = \frac{3}{2}{\cal A}(D^+ \pi^-) - \frac{1}{2}{\cal A}(D^0 \pi^-)
= \frac{1}{2} C - T - \frac{3}{2} E~~~,
\eeq
with similar relations for $D^* \pi$ and $D \rho$ decays.
The corresponding relation for Cabibbo-suppressed decays is
\beq
A_1^{DK} = {\cal A}(D^0 K^-) = -(T'+C') ~, ~~~
A_0^{DK} = {\cal A}(D^+ K^-) - {\cal A}(D^0 \overline{K}^0) = C'-T' ~,
\eeq
with similar relations for $D^* K$ and $D K^*$ decays.

\bigskip
\centerline{\bf IV.  TOPOLOGICAL AMPLITUDES:  MAGNITUDES AND PHASES}
\bigskip

In Tables I--III we summarize the rates, invariant amplitudes, and their
flavor-SU(3) representations for decays of $\overline{B}$ mesons to $D \pi$,
$D^* \pi$, and $D \rho$, respectively.  Also shown are decays to other final
states related by flavor SU(3).  Branching ratios and lifetimes are taken from
the compilation of Ref.~\cite{PDG} except where indicated otherwise.  In
particular, we take $\tau(B^-) = (1.674 \pm 0.018) \times 10^{-12}$ s,
$\tau(\ob) = (1.542 \pm 0.016) \times 10^{-12}$ s.  For the $D \pi$ decays we
use the updated values quoted in Refs.\ \cite{PedlarD,Ahmed}.  The branching
ratio for $\ob \to D_s^+ K^-$ is based on our average of new values from Belle
\cite{BeDsK}:  ${\cal B} = (4.6^{+1.2}_{-1.1} \pm 1.3) \times 10^{-5}$ and
BaBar \cite{BaDsK}: ${\cal B} = (3.2 \pm 1.0 \pm 1.0) \times 10^{-5}$.
The branching ratios and limits for $\ob \to D^{(*)0} \overline{K}^{(*)0}$
are based on a recent report by the Belle Collaboration \cite{BeDK}.

In Table I the amplitudes $T$, $C$, and $E$ were described above; in Tables II
and III the amplitudes are labelled with subscripts which denote the meson
containing the spectator quark: $P$ for pseudoscalar, $V$ for vector
\cite{DGR}.  We omit contributions of disconnected diagrams \cite{ChauDisc,BFT}
in which $\eta$ and $\eta'$ exchange no quark lines with the rest of the
diagram, and couple through their SU(3)-singlet components.

Tables I--III contain several tests of flavor SU(3).  The breaking of this
symmetry is incorporated via ratios of decay constants: $f_K/f_\pi = 1.22$,
$f_{K^*}/f_\rho = 1.04$.  For example, the decay $B^- \to D^0 K^-$ is related
to $D^- \to D^0 \pi^-$ by the U-spin substitution $d \to s$, and so one expects
${\cal A}(B^- \to D^0 K^-)/{\cal A}(D^- \to D^0 \pi^-) = (f_K/f_\pi)(\lambda/
[1-\frac{\lambda^2}{2}])=0.275$, where $\lambda = 0.22$ describes the hierarchy
of CKM matrix elements \cite{WP}
and the small form factor difference is ignored throughout the paper.
When one corrects for this factor, the
derived values of $|T+C|$ are equal within errors.  Similar results hold for
the ratio ${\cal A}(B^- \to D^{*0} K^-)/{\cal A}(D^- \to D^{*0} \pi^-) =
(f_K/f_\pi)(\lambda/[1 - \frac{\lambda^2}{2}]) = 0.275$ and ${\cal A}(B^- \to
D^0 K^{*-})/{\cal A}(D^- \to D^0 \rho^-) = (f_{K*}/f_\rho)(\lambda/[1 -
\frac{\lambda^2}{2}]) = 0.235$.

The amplitudes for $\ob \to D^+ \pi^-$ and $\ob \to D^+ K^-$ (and similar modes
with one final-state pseudoscalar replaced by a vector meson) would be related
to one another by U-spin if one neglected the presence of the spectator quark.
The spectator quark contributes an additional exchange amplitude, whose
magnitude is seen to be small from the decay $\bo \to D_s^+ K^-$ in Table I and
the upper limit on $\bo \to D_s^{*+} K^-$ in Table II.  Thus, for example, one
cannot tell the difference between $|T+E|$ extracted from $\ob \to D^+ \pi^-$
and $|T|$ extracted from $\ob \to D^+ K^-$.  A similar conclusion applies to
$|T_V+E_P|$ versus $|T_V|$ in Table II and $|T_P+E_V|$ versus $|T_P|$ in Table
III.

\begin{table}
\caption{Rates and invariant amplitudes for decays of $\overline{B}$ mesons
mesons to $D \pi$ and related modes. Primed amplitudes are related to unprimed
amplitudes by a factor of $\lambda
f_K/\left[ f_\pi(1-\frac{\lambda^2}{2}) \right] = 0.275$.
Except where noted, the branching ratios are quoted from the Particle Data
Group \cite{PDG}.}
\begin{center}
\begin{tabular}{r c c c c c} \hline \hline
Decay & $M$   & Branching ratio & $p^*$ & $|{\cal A}|$ & Representation \\
      & (GeV) & (units of $10^{-4}$) & (GeV) & $(10^{-7} {\rm GeV})$ \\
\hline
$B^- \to D^0 \pi^-$ & 5.2790 & $49.7 \pm 3.8^{~a} $ & 2.308 & $7.70 \pm 0.29$
                        & $-(T+C)$ \\
$    \to D^0 K^-$   &        & $3.7 \pm 0.6$ & 2.281 & $2.11 \pm 0.17$
                        & $-(T'+C')$ \\
     & & & & $7.7 \pm 0.6^{~b}$ & $|T+C|$ \\ \hline
$\ob \to D^+ \pi^-$ & 5.2794 & $26.8 \pm 2.9^{~a}$ & 2.306 & $5.89 \pm 0.32$
                        & $-(T+E) $ \\
$   \to  D^+ K^-$   &        & $2.0 \pm 0.6$ & 2.279 & $1.62 \pm 0.24$
                        & $-T'$ \\
     & & & & $5.9 \pm 0.9^{~b}$ & $|T|$ \\ 
$    \to D^0 \pi^0$ &        & $2.92 \pm 0.45^{~a}$ & 2.308 & $1.94 \pm 0.15$
                        & $(E-C)/\s$ \\
$   \to  D^0 \eta$  &   & $1.4^{+0.6}_{-0.5}$ & 2.274 & $1.36 \pm 0.27$
                        & $(C+E)/\st$ \\
$   \to  D^0 \eta'$ &        &     $ < 9.4 $  & 2.198 & $ < 3.6 $
                        & $-(C+E)/\sx$ \\
$   \to D^0 \ok$ & & $0.50^{+0.13}_{-0.12} \pm 0.06$ & 2.280 & $0.81 \pm 0.11$
                        & $-C'$ \\
     & & & & $2.94 \pm 0.41^{~b}$ & $|C|$ \\
$   \to  D_s^+ K^-$ & & $0.38 \pm 0.10 \,^d$ & 2.242 & $0.71 \pm 0.10$
                        & $-E$ \\
\hline \hline
\end{tabular}
\end{center}
\leftline{$^a$ Refs.\ \cite{PedlarD,Ahmed}.  $^b$ Value implied by (broken)
flavor SU(3).  $^c$ Ref.\ \cite{BeDK}.  $^d$ Avg.\ of \cite{BeDsK,BaDsK}.}
\end{table}

\begin{table}
\caption{Rates and invariant amplitudes for decays of $\overline{B}$ mesons
mesons to $D^* \pi$ and related modes.  Primed amplitudes are related to
unprimed amplitudes by a factor of $\lambda f_K/\left[ f_\pi(1 -
  \frac{\lambda^2}{2}) \right] = 0.275$.
The branching ratios are quoted from the Particle Data Group \cite{PDG}.}
\begin{center}
\begin{tabular}{r c c c c c} \hline \hline
Decay & $M$   & Branching ratio & $p^*$ & $|{\cal A}|$ & Representation \\
      & (GeV) & (units of $10^{-4}$) & (GeV) & $(10^{-7})$ \\
\hline
$B^- \to D^{*0} \pi^-$ & 5.2790 & $46 \pm 4 $ & 2.256 & $3.32 \pm 0.14$
                        & $-(T_V+C_P)$ \\
$    \to D^{*0} K^-$   &        & $3.6 \pm 1.0$ & 2.227 & $0.95 \pm 0.13$
                        & $-(T'_V+C'_P)$ \\
     & & & & $3.4 \pm 0.5^{~a}$ & $|T_V + C_P|$ \\ \hline
$\ob \to D^{*+} \pi^-$ & 5.2794 & $27.6 \pm 2.1$ & 2.255 & $2.68 \pm 0.10$
                        & $-(T_V+E_P) $ \\
$   \to  D^{*+} K^-$   &        & $2.0 \pm 0.5$ & 2.226 & $0.74 \pm 0.09$
                        & $-T_V'$ \\
     & & & & $2.7 \pm 0.3^{~a}$ & $|T_V|$ \\
$    \to D^{*0} \pi^0$ &        & $2.5 \pm 0.7$ & 2.256 & $0.81 \pm 0.11$
                        & $(E_P-C_P)/\s$ \\
$   \to  D^{*0} \eta$  &   & $ < 2.6 $ & 2.220 & $ < 0.84 $
                        & $(C_P+E_P)/\st$ \\
$   \to  D^{*0} \eta'$ &        &     $ < 14 $  & 2.141 & $ < 2.1 $
                        & $-(C_P+E_P)/\sx$ \\
$   \to D^{*0} \ok$    & & $ < 0.66^{~b}$ & 2.227 & $ < 0.42 $
                        & $-C_P'$ \\
     & & & & $ < 1.54^{~a}$ & $|C_P|$ \\
$   \to  D_s^{*+} K^-$ & & $ < 0.25 $ & 2.185 & $ < 0.27 $
                        & $-E_P$ \\
\hline \hline
\end{tabular}
\end{center}
\leftline{$^a$ Value implied by (broken) flavor SU(3).  $^b$ Ref.\
\cite{BeDK}.}
\end{table}

\begin{table}
\caption{Rates and invariant amplitudes for decays of $\overline{B}$ mesons
mesons to $D \rho$ and related modes.  Primed amplitudes are related to
unprimed amplitudes by a factor of $\lambda f_{K*}/\left[ f_\rho (1 -
  \frac{\lambda^2}{2}) \right] = 0.235$.
Except where noted, the branching ratios are quoted from the Particle Data
Group \cite{PDG}.}
\begin{center}
\begin{tabular}{r c c c c c} \hline \hline
Decay & $M$   & Branching ratio & $p^*$ & $|{\cal A}|$ & Representation \\
      & (GeV) & (units of $10^{-4}$) & (GeV) & $(10^{-7})$ \\
\hline
$B^- \to D^0 \rho^-$ & 5.2790 & $134 \pm 18 $ & 2.238 & $5.74 \pm 0.39$
                        & $-(T_P+C_V)$ \\
$    \to D^0 K^{*-}$   &        & $6.1 \pm 2.3$ & 2.213 & $1.25 \pm 0.23$
                        & $-(T'_P+C'_V)$ \\
       & & & & $5.3 \pm 1.0^{~a}$ & $|T_P + C_V|$ \\ \hline
$\ob \to D^+ \rho^-$ & 5.2794 & $78 \pm 14$ & 2.236 & $4.57 \pm 0.41$
                        & $-(T_P+E_V) $ \\
$   \to  D^+ K^{*-}$   &        & $3.7 \pm 1.8$ & 2.211 & $1.01 \pm 0.25$
                        & $-T'_P$ \\
       & & & & $4.3 \pm 1.0^{~a}$ & $ |T_P|$ \\
$\to D^0 \rho^0$ &      & $2.9 \pm 1.0 \pm 0.4 \,^b$ & 2.238 & $0.88 \pm 0.16$
                        & $(E_V-C_V)/\s$ \\
$   \to  D^0 \omega$  &   & $1.8 \pm 0.6$ & 2.235 & $0.69 \pm 0.12$
                        & $-(C_V+E_V)/\s$ \\
$\to D^0 \overline{K}^{*0}$ & & $0.48^{+0.11}_{-0.10} \pm 0.05^{~c}$ & 2.212 &
   $0.36 \pm 0.04$ & $-C_V'$ \\
       & & & & $1.55 \pm 0.19^{~a}$ & $|C_V|$ \\
$   \to  D_s^+ K^{*-}$ & & $ < 9.9 $ & 2.172 & $ < 1.7$
                        & $-E_V$ \\
\hline \hline
\end{tabular}
\end{center}
\leftline{$^a$ Value implied by (broken) flavor SU(3). 
$^b$ Ref.\ \cite{BelDrho}.  $^c$ Ref.\ \cite{BeDK}.}
\end{table}

We now discuss the amplitude triangles for each set of processes.  It is
interesting to determine the individual magnitudes and phases of the
contributing topological amplitudes $T$, $C$, and $E$, as was done for charmed
particle decays \cite{JRcharm}.  This can be important for understanding the
systematics of $B$ decays involving two amplitudes with different weak {\it
  and} strong phases, for which direct CP violation can be observed.  In the
present case, of course, it is only the strong phases which may differ from one
another.  The decays $\overline{B} \to D \pi$ and those related to it by flavor
SU(3) permit one to map out the amplitudes (up to a discrete ambiguity), while
those involving one vector meson in the final state are missing key information
which one hopes will be provided by BaBar, Belle, or hadron colliders.
\bigskip

\leftline{\bf A.  $\overline{B} \to D \pi$ and related decays}
\bigskip

The amplitudes $A(B^- \to D^0 \pi^-) = - (T+C)$, $A(\ob \to D^+ \pi^-) = -(T
+ E)$, and $\sqrt{2} A(\ob \to D^0 \pi^0) = E-C$ form a triangle in the complex
plane, as shown in Fig.\ 1.  Here we have arbitrarily taken $T+C$ to be real
and positive. The favored area of the triangle is non-zero, as our earlier
discussion of isospin amplitudes also implies.
 
The decay $\ob \to D_s^+ K^-$ \cite{BaDsK} provides a value of $|E|$, whose
central value is used to draw a circle of radius $|E|$ around the intersection
of the two sides $T+E$ and $C-E$.  The decay $\ob \to D^0 \eta$ provides a
value of $|C+E|$.  Using the relation $(C-E)/2 + E = (C+E)/2$ for complex
amplitudes, we draw a circle of radius $|C+E|/2$ about the midpoint of the side
$C-E$.  The intersections $O$ and $O'$ of the two circles then denote the
allowed phases of $E$.  One can now identify the amplitudes $T$ and $C$
corresponding to each of these solutions.

\begin{figure}
\begin{center}
\includegraphics[height=2.1in]{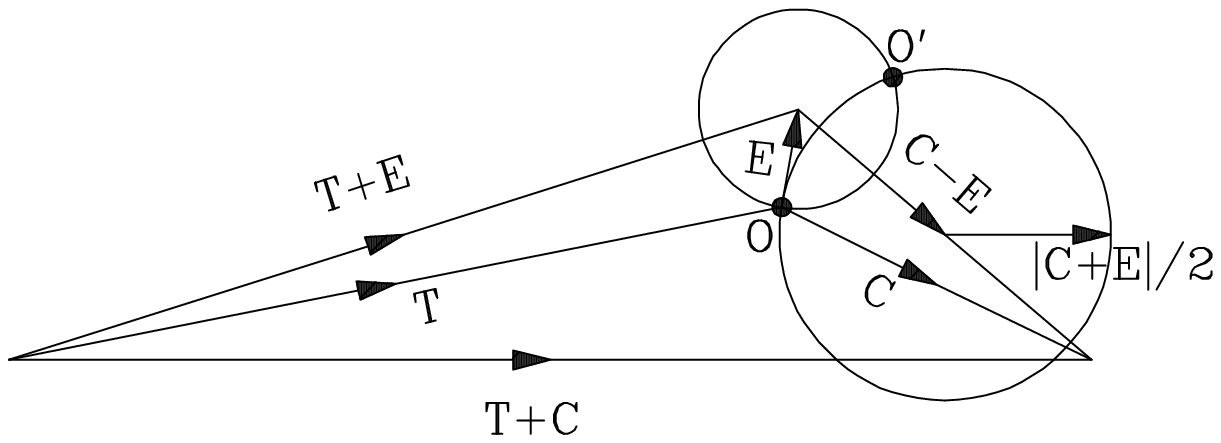}
\caption{Amplitude triangle for $\overline{B} \to D \pi$ and related decays.
The amplitude $E$ points from either $O$ or $O'$ to the center
of the small circle.  The amplitudes $T$ and $C$ are shown only for the first
of these two solutions.}
\end{center}
\end{figure}

In principle the value of $|T|$, provided through broken flavor SU(3) by the
decay $\ob \to D^+ K^-$, could help to resolve the discrete ambiguity.  In the
solution shown in Fig.\ 1, we have $|T| \simeq 5.6$ (here and in the following
analysis, we express topological amplitudes in units of $10^{-7}$ GeV for $PP$
modes and $10^{-7}$ for $PV$ modes), while in the solution in which $T$ points
to $O'$, one has $|T| \simeq 6.6$.  The error on $|T| = 5.9 \pm 0.9$ from $\ob
\to D^+ K^-$ is at least a factor of three too large to permit any distinction
between the two solutions.

Similarly, the value $|C| = 2.94 \pm 0.41$, obtained via flavor SU(3) from
the recently reported decay \cite{BeDK} $\ob \to D^0 \ok$, can be compared
with that implied by Fig.\ 1, which is the same for the two discrete solutions.
The measurements in Table I provide $|C \pm E|$ and $|E|$.  One can then solve
to find $|C| = 2.46 \pm 0.25$, consistent with the above value.

The solution shown has some similarity to that for $D \to \overline{K} \pi$
\cite{JRcharm}.  In that solution, the amplitudes $T$, $C$, and $E$ all had
distinctly different phases.  Denoting $\delta_{AB}$ as the angle of rotation
from the amplitude $B$ to $A$, then we have as the central values $\delta_{CT}
\simeq -38^{\circ}$ and $\delta_{ET} \simeq 69^{\circ}$.  The other solution
(not shown) has $T$ and $E$ relatively real, both with a large phase relative
to $C$.  Explicitly, we obtain $\delta_{CT} \simeq -73^{\circ}$ and
$\delta_{ET} \simeq 180^{\circ}$.  In comparison, the phases in the
corresponding $D$ decays are $\delta_{CT} \simeq -151^{\circ}$ and $\delta_{ET}
\simeq 115^{\circ}$ \cite{JRcharm}.
The sign flip in Re$(C/T)$ (negative for charm, positive for beauty) also
occurs in a simplified analysis in which $C/T$ is taken to be real and the
effects of the amplitude $E$ are not taken into account.  In such a case the
sign flip is merely a consequence of $\Gamma(D^+ \to \ok \pi^+) < \Gamma(D^0
\to K^- \pi^+)$, $\Gamma(B^- \to D^0 \pi^-) > \Gamma(\ob \to D^+ \pi^-)$.

It is interesting to compare the ratios $|C/T|$ and $|E/T|$ with those found
for $D \to \overline{K} \pi$, where our favored solution \cite{JRcharm}
had $|C/T| \simeq 0.8$ and $|E/T| \simeq 0.6$.  Here, the solution shown in
Fig.\ 1 has $|C/T| \simeq 0.4$ and $|E/T| \simeq 0.1$.  The amplitude $E$ is
indeed vanishing faster than the others as the heavy quark mass $m_Q$
increases, in accord with expectations for heavy-quark systems, but as some
power between $m_Q^{-1}$ and $m_Q^{-2}$.

We comment briefly on the claim by Xing \cite{Xing} that the pattern of
branching ratios in $\overline{B} \to D \overline{K}$ decays implies non-zero
final-state phases between isospin amplitudes.  This is certainly true for
central values.  However, the sum rule (\ref{eqn:sr}) when written for
amplitudes which are relatively real reads
\beq
(2.43 \pm 0.27) \times 10^{-7} {\rm~GeV} =
(2.11 \pm 0.17) \times 10^{-7} {\rm~GeV}
\eeq
using the amplitudes in Table I.  This is satisfied at the $1 \sigma$ level.
The errors are dominated by those for the color-favored processes; reduction
by a factor of 2 would help considerably.

\bigskip
\leftline{\bf B.  $\overline{B} \to D^* \pi$ and related decays}
\bigskip

The triangle formed by the amplitudes $A(B^- \to D^{*0} \pi^-) = - (T_V+C_P)$,
$A(\ob \to D^{*+} \pi^-) = -(T_V + E_P)$, and $\sqrt{2} A(\ob \to D^{*0} \pi^0)
= E_P-C_P$ is shown in Fig.\ 2.  We have taken $T_V+C_P$ to be real and
positive.  Non-zero area again is favored.

\begin{figure}
\begin{center}
\includegraphics[height=2.4in]{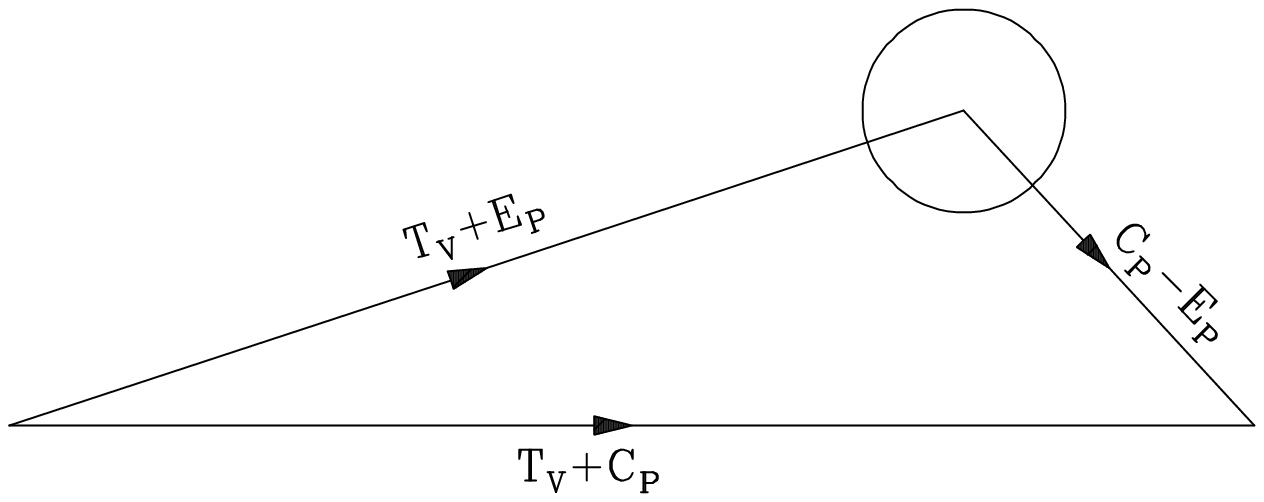}
\caption{Amplitude triangle for $\overline{B} \to D^* \pi$ and related decays.
The amplitude $E_P$ points from anywhere inside the small circle to the
intersection of the lines $T_V+E_P$ and $T_V+C_P$.}
\end{center}
\end{figure}

Here the situation is much less satisfactory than for $\overline{B} \to D \pi$.
We have only an upper bound on $|E_P|$ based on the non-observation of $\ob \to
D_s^{*+} K^-$.  The decay $\ob \to D^{*+} K^-$ tells us that $|T_V| = 2.7 \pm
0.3$, whereas on the basis of $|E_P| \le 0.27$ and $|T_V+E_P| = 2.68 \pm 0.10$
(Table II) we could have any value of $|T_V|$ between 2.2 and 3.1.  We also
have no information on $|C_P+E_P|$, only a poor upper bound from $\ob \to
D^{*0} \eta$ and a much worse one from $\ob \to D^{*0} \eta'$.  If the pattern
in Fig.\ 2 is anything like that for the corresponding charm decays $D \to
\overline{K}^* \pi$, it should resemble that for the solution displayed in
Fig.\ 1.  Improved information on $|C_P|$, obtainable via flavor SU(3) from the
decay $\ob \to D^{*0} \ok$ for which only an upper limit is quoted \cite{BeDK},
would also be helpful.  This process also would be useful in implementing the
suggestion of Xing \cite{Xing} to search for relative final-state phases
between isospin amplitudes in $\overline{B} \to D^* \overline{K}$.
\newpage

\leftline{\bf C.  $\overline{B} \to D \rho$ and related decays}
\bigskip

The triangle formed by the amplitudes $A(B^- \to D^0 \rho^-) = - (T_P+C_V)$,
$A(\ob \to D^+ \rho^-) = -(T_P + E_V)$, and $\sqrt{2} A(\ob \to D^0 \rho^0) =
E_V-C_V$, shown in Fig.\ 3, has a much smaller area than either of the previous
two, and is consistent with the same phase for each of the three amplitudes.
Here we have taken $T_P+C_V$ to be real and positive.

\begin{figure}
\begin{center}
\includegraphics[height=1.3in]{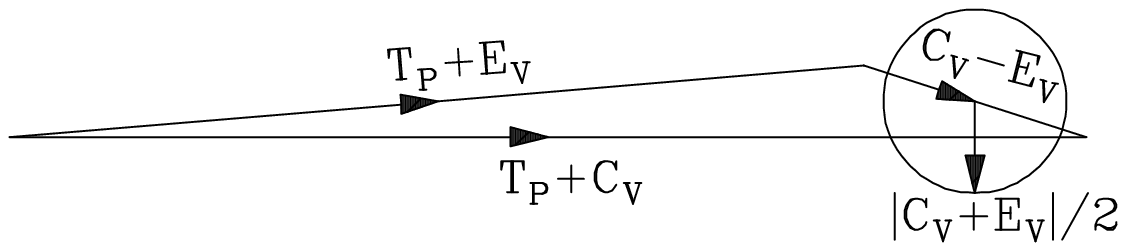}
\caption{Amplitude triangle for $\overline{B} \to D \rho$ and related decays.}
\end{center}
\end{figure}

Additional information is available from the decay $\ob \to D^0 \omega$,
which provides a value of $|C_V+E_V|$.  As in Fig.\ 1, we draw a circle
of radius $|C_V+E_V|/2$ from the midpoint of the line $C_V-E_V$.  The solution
points $O$ and $O'$ would correspond to the intersection of this circle with
one of radius $|E_V|$ whose center is the intersection of the lines $T_P + E_V$
and $C_V-E_V$.  We would need an improved upper bound on $\ob \to D_s^+ K^{*-}$
in order to draw a useful version of this last circle.

As in the two previous cases, an estimate of the tree amplitude ($|T_P|$ in
this case) would also be helpful.  From the decay $\ob \to D^0 K^{*-}$ we find
$|T_P| = 4.3 \pm 1.0$, to be compared with $|T_P + E_V| = 4.57 \pm 0.41$.
Obviously no conclusion can be drawn at present about the relative phase of
$T_P$ and $E_V$.

Flavor SU(3) can be applied to the decay $\ob \to D^0 \overline{K}^{*0}$,
yielding the magnitude $|C_V| = 1.55 \pm 0.19$ quoted in Table III.  An
independent upper limit on this quantity can be obtained by combining
information on $|C_V + E_V|$ from $\ob \to D^0 \rho^0$ and $|C_V - E_V|$
from $\ob \to D^0 \omega$ to obtain $(|C_V|^2 + |E_V|^2)^{1/2} = 1.12 \pm
0.15$.  There is clearly not very much room for $|E_V|$ at the upper limit
quoted in Table III, and even the hint ($< 2 \sigma$) of an inconsistency.
The most incisive test would probably be direct observation of the decay
$\ob \to D_s^+ K^{*-}$, providing a value of $|E_V|$.

As in the case of $\overline{B} \to D \overline{K}$, the decays $\overline{B}
\to D \overline{K}^*$ in principle provide information on relative strong
final-state phases \cite{Xing}.  Here the sum rule (\ref{eqn:sr}), if
written for relatively real amplitudes, would read
\beq
(1.38 \pm 0.25) \times 10^{-7} = (1.25 \pm 0.23) \times 10^{-7}
\eeq
using the amplitudes in Table III.  This is satisfied at considerably better
than $1 \sigma$.  Again, the error is dominated by that in the color-favored
processes.

If the pattern of $\overline{B} \to D \rho$ decays is similar to that for the
corresponding $D \to
\overline{K} \rho$ decays, the triangle in Fig.\ 3 would assume its squashed
shape as a result of a negative largely imaginary contribution of $E_V$, so
that the relative phase of $T_P$ and $C_V$ would be non-trivial.  In the case
of $\overline{B} \to D^* \pi$ decays (Fig.\ 2), by contrast, the contribution
of $E_P$ would be largely imaginary and positive, leading to a triangle with
greater area as in the case of $\overline{B} \to D \pi$ (Fig.\ 1).  This
pattern is what occurs in charm decays \cite{JRcharm}.  We now discuss what
measurements might determine if a similar picture applies to $\overline{B}$
decays.
\bigskip

\centerline{\bf V.  MISSING PIECES OF THE PUZZLE}
\bigskip

We began by asking the question of whether the pattern of $T$, $C$, and $E$
amplitudes in the decays $\overline{B} \to D \pi,~D^* \pi$, and $D \rho$ bore
any relation to that in $D \to \overline{K} \pi,~\overline{K}^* \pi$, and
$\overline{K} \rho$.  We see that there is some resemblance of the two cases in
that the decay amplitudes for the first two processes appear to have
non-trivial relative phases which could well be absent for each of the third
processes.  However, we are frustrated in our quest for the topological
amplitudes by the fragmentary nature of the data.  We believe this situation
could well improve in the near future.

One key element responsible for the pattern in charm decays is the flip of the
sign of the exchange amplitude when one interchanges which particle in the
final state is a pseudoscalar and which is a vector.  Thus, for charm, we found
that the relation $E_V = - E_P$, which could be justified if the exchange
amplitude really proceeded through a quark-antiquark state \cite{HJL}, was
responsible for the very different pattern of amplitudes in $D \to
\overline{K}^* \pi$ and $D \to \overline{K} \rho$.  We do not yet have enough
information to draw such a conclusion for $\overline{B} \to D^* \pi$ and
$\overline{B} \to D \rho$ decays.

In Table IV we summarize some useful decays that would help to sort out the
question of the topological amplitudes.  We quote in each case the present
error and a desirable error in ${\cal B}$, in units of $10^{-5}$.  We see that
improvements by factors of three in branching ratios or roughly a ten-fold
increase in the data sample would permit a fairly clear pattern to emerge.
There may be some short-cuts to this procedure which would be less demanding in
data.  We now briefly justify each of the entries in Table IV.

\begin{table}
\caption{$\overline{B}$ decays which would provide useful information on
topological amplitudes for $\overline{B} \to D \pi,~D^* \pi$, and $D \rho$.}
\begin{center}
\begin{tabular}{c c c c} \hline \hline
Amplitude & Decay & Present & Desirable \\
          &      & \multicolumn{2}{c}{Error in ${\cal B} \times 10^5$}\\ \hline
$|T|$ & $\ob \to D^+ K^-$ & 6 & 1.3 \\
$|T+C|$ & $B^- \to D^0 \pi^-$ & 38 & 26 \\
$|T+C|$ & $B^- \to D^0 K^-$ & 6 & 2 \\
$|C|$ & $\ob \to D^0 \ok$ & 1.4 & 0.7 \\
$|C+E|$ & $\ob \to D^0 \eta$ & 6 & 2 \\
$|T_V|$ & $\ob \to D^{*+} K^-$ & 5 & 1.5  \\
$|C_P|$ & $\ob \to D^{*0} \ok$ & $<6.6^{~a}$ & 0.7 \\
$|C_P + E_P|$ & $\ob \to D^{*0} \eta$ & $<26^{~a}$ & 2 \\
$|E_P|$ & $\ob \to D_s^{*+} K^-$ & $<2.5^{~a}$ & 1 \\
$|T_P|$ & $\ob \to D^+ K^{*-}$ & 18 & 2 \\
$|T_P + C_V|$ & $B^- \to D^0 \rho^-$ & 180 & 45 \\
$|T_P + C_V|$ & $B^- \to D^0 K^{*-}$ & 23 & 2 \\
$|E_V|$ & $\ob \to D_s^+ K^{*-}$ & $<99^{~a}$ & 1 \\
\hline \hline
\end{tabular}
\end{center}
\leftline{$^a$ Present 90\% c.l.\ upper limit on branching ratio.}
\end{table}

It would be helpful to have the error on $|T|$ from the decay $\ob \to D^+ K^-$
small enough that one could tell the difference between $|T+E|$ and $|T|$ at
least for the case of maximal constructive or destructive $T$--$E$
interference.  Thus, we ask for the error on $|T|$ to be less than 1/3 the
value of $|E|$, or $\Delta |T| \le 0.2$.  (Our convention for units was
mentioned in Sec. IV A.)  This is a factor of $4.5$ increase in present
accuracy, both in ${\cal A}$ and in branching ratio.
We also require $\Delta |T+C| \le 0.2$ as obtained from $\ob \to D^0 \pi^-$
and, therefore, the error on the branching ratio should be reduced by roughly a
factor of $1.5$.  We ask for similar errors in
$|T+C|$ and $|C|$ as obtained via flavor SU(3) from $B^- \to D^0 K^-$ and $\ob
\to D^0 \ok$, respectively, and in $|C+E|$ obtained from $\ob \to D^0 \eta$.
Demanding that $\Delta |T+C| = \Delta |C| = \Delta |C+E| \le 0.2$ we find that
the corresponding branching ratios should be specified to an error of $\pm
(2,~0.7,~2) \times 10^{-5}$.  The proposed errors on $|T|$, $|T+C|$, and $|C|$
also would allow one to draw a useful conclusion about the relative strong
phases of isospin amplitudes in $\overline{B} \to D \overline{K}$.

If $\ob \to D^{*+} K^-$ is to provide a useful value of $|T_V|$ to compare
with $|T_V+E_P|$ from $\ob \to D^+ \pi^-$, we need $\Delta |T_V|$ to be
no more than 1/3 of $|E_P|$, or at most $0.1$.  This, again,
represents a three-fold improvement in the error on the branching ratio.
The color-suppressed decay $\ob \to D^{*0} \ok$ would provide a useful
value of $|C_P|$ via flavor SU(3) if measured with an error similar to
that desired for $\ob \to D^0 \ok$, or $\Delta {\cal B} = 0.7 \times 10^{-5}$.

The decay $\ob \to D^{*0} \eta$ is very poorly measured, corresponding only
to a rather weak upper bound.  An error on its branching ratio comparable to
that for $\ob \to D^0 \eta$ would be highly desirable.

Evidence for the exchange amplitude $E_P$ at something approaching the $3
\sigma$ level would be useful for the present program.  Thus, if the branching
ratio for $\ob \to D^{*+}_s K^-$ is close to its present upper limit
of $2.5 \times 10^{-5}$, an error of no more than $1 \times 10^{-5}$ would be
desirable.

The likelihood that $E_V = - E_P$ sets the scale of useful errors in decays
related to $\overline{B} \to D \rho$.  One would like errors in $|T_P|$
(from $\ob \to D^+ K^{*-}$) and $|E_V|$ (from $\ob \to D_s^+ K^{*-}$) no
larger than $0.1$, leading to the rather stringent demands
in Table IV.
A similar error requirement on $|T_P + C_V|$ as obtained directly from $B^- \to
D^0 \rho^-$ means only a four-fold improvement in the branching ratio.
It may be worth searching for alternative strategies to
constrain these amplitudes.
The proposed error on $B^- \to D^0 K^{*-}$, when combined with other proposed
measurements, will be more than sufficient to check the relative phases of
isospin amplitudes in $\overline{B} \to D \overline{K}^*$.

Effects of the exchange amplitudes $E,~E_P$, and $E_V$ (indeed, also of
$C,~C_P$, and $C_V$) may be generated by rescattering effects
\cite{JRcharm,NP,BGR,HYC}, as has been emphasized in reports of the decay $\ob
\to D_s^+ K^-$ \cite{BeDsK,BaDsK}.  In this case we may not be able to justify
the assumption \cite{HJL} $E_V = -E_P$.  However, this relation does appear
consistent with charm decays \cite{JRcharm}, and for the moment with beauty
decays as well.

\bigskip
\centerline{\bf VI.  SUMMARY AND DISCUSSION}
\bigskip

We have compared the decays $\overline{B} \to D \pi,~D^* \pi$, and $D \rho$
with the corresponding charmed particle decays $D \to \overline{K} \pi,
\overline{K}^* \pi$, and $\overline{K} \rho$.  In the first two of each set,
there appear to be non-trivial final-state phases between the decay amplitudes,
while in the third case in each set, the decay amplitudes appear to be
relatively real.

In the case of charm decays, we traced the apparent relative reality of $D \to
\overline{K} \rho$ amplitudes to an accidental cancellation of non-trivial
final-state phases among the tree ($T$), color-suppressed ($C$), and exchange
($E$) amplitudes.  Our analysis of $\overline{B}$ decays suggests that while
the exchange amplitude is diminishing in importance, with $|E/T| \simeq 0.1$
for $\overline{B} \to D \pi$ as compared with about 0.6 for $D \to \overline{K}
\pi$, it still can play a significant role in contributing to the observed
final-state phases, at least for $\overline{B} \to D \pi$.  We have identified
a number of measurements which could determine whether the apparently different
shapes of the amplitude triangles for $\overline{B} \to D^* \pi$ (Fig.\ 2) and
$\overline{B} \to D \rho$ (Fig.\ 3) are due to a simple sign flip of the
exchange amplitude, as occurs for charm.  We have also indicated improvements
in accuracy likely to be needed to identify non-zero final-state phases between
amplitudes in $\overline B \to D \overline{K}$ and related decays.

\bigskip
\centerline{\bf ACKNOWLEDGEMENTS}
\bigskip

We thank A. Bondar, S. Olsen, A. A. Petrov, V. Sharma, and S. F. Tuan for
helpful correspondence. This work was supported in part by the United States
Department of Energy, High Energy Physics Division, under Contract Nos.\
DE-FG02-90ER-40560 and W-31-109-ENG-38. 
\bigskip

\def \ajp#1#2#3{Am.~J.~Phys.~{\bf#1}, #2 (#3)}
\def \apny#1#2#3{Ann.~Phys.~(N.Y.) {\bf#1}, #2 (#3)}
\def \app#1#2#3{Acta Phys.~Polonica {\bf#1}, #2 (#3)}
\def \arnps#1#2#3{Ann.~Rev.~Nucl.~Part.~Sci.~{\bf#1}, #2 (#3)}
\def \art{and references therein}
\def \b97{{\it Beauty '97}, Proceedings of the Fifth International
Workshop on $B$-Physics at Hadron Machines, Los Angeles, October 13--17,
1997, edited by P. Schlein}
\def \carg{{\it Masses of Fundamental Particles -- Carg\`ese 1996}, edited by
M. L\'evy \ite, NATO ASI Series B:  Physics Vol.~363 (Plenum, New York, 1997)}
\def \cmp#1#2#3{Commun.~Math.~Phys.~{\bf#1}, #2 (#3)}
\def \cmts#1#2#3{Comments on Nucl.~Part.~Phys.~{\bf#1}, #2 (#3)}
\def \corn93{{\it Lepton and Photon Interactions:  XVI International
Symposium, Ithaca, NY August 1993}, AIP Conference Proceedings No.~302,
ed.~by P. Drell and D. Rubin (AIP, New York, 1994)}
\def \cp89{{\it CP Violation,} edited by C. Jarlskog (World Scientific,
Singapore, 1989)}
\def \dpff{{\it The Fermilab Meeting -- DPF 92} (7th Meeting of the
American Physical Society Division of Particles and Fields), 10--14
November 1992, ed. by C. H. Albright \ite~(World Scientific, Singapore,
1993)}
\def \dpf94{DPF 94 Meeting, Albuquerque, NM, Aug.~2--6, 1994}
\def \efi{Enrico Fermi Institute Report No. EFI}
\def \el#1#2#3{Europhys.~Lett.~{\bf#1}, #2 (#3)}
\def \epjc#1#2#3{Eur.~Phys.~J.~C {\bf#1}, #2 (#3)}
\def \f79{{\it Proceedings of the 1979 International Symposium on Lepton
and Photon Interactions at High Energies,} Fermilab, August 23-29, 1979,
ed.~by T. B. W. Kirk and H. D. I. Abarbanel (Fermi National Accelerator
Laboratory, Batavia, IL, 1979}
\def \hb87{{\it Proceeding of the 1987 International Symposium on Lepton
and Photon Interactions at High Energies,} Hamburg, 1987, ed.~by W. Bartel
and R. R\"uckl (Nucl. Phys. B, Proc. Suppl., vol. 3) (North-Holland,
Amsterdam, 1988)}
\def \ib{{\it ibid.}~}
\def \ibj#1#2#3{{\it ibid.}~{\bf#1}, #2 (#3)}
\def \ichep72{{\it Proceedings of the XVI International Conference on High
Energy Physics}, Chicago and Batavia, Illinois, Sept. 6--13, 1972,
edited by J. D. Jackson, A. Roberts, and R. Donaldson (Fermilab, Batavia,
IL, 1972)}
\def \ijmpa#1#2#3{Int.~J.~Mod.~Phys.~A {\bf#1}, #2 (#3)}
\def \ite{{\it et al.}}
\def \jmp#1#2#3{J.~Math.~Phys.~{\bf#1}, #2 (#3)}
\def \jpg#1#2#3{J.~Phys.~G {\bf#1}, #2 (#3)}
\def \lkl87{{\it Selected Topics in Electroweak Interactions} (Proceedings
of the Second Lake Louise Institute on New Frontiers in Particle Physics,
15--21 February, 1987), edited by J. M. Cameron \ite~(World Scientific,
Singapore, 1987)}
\def \KEK#1{{\it Flavor Physics} (Proceedings of the Fourth International
Conference on Flavor Physics, KEK, Tsukuba, Japan, 29--31 October 1996),
edited by Y. Kuno and M. M. Nojiri, Nucl.~Phys.~B Proc.~Suppl.~{\bf 59},
#1 (1997)}
\def \ky85{{\it Proceedings of the International Symposium on Lepton and
Photon Interactions at High Energy,} Kyoto, Aug.~19-24, 1985, edited by M.
Konuma and K. Takahashi (Kyoto Univ., Kyoto, 1985)}
\def \mpla#1#2#3{Mod.~Phys.~Lett.~A {\bf#1}, #2 (#3)}
\def \nc#1#2#3{Nuovo Cim.~{\bf#1}, #2 (#3)}
\def \nima#1#2#3{Nucl.~Instr.~Meth.~A {\bf#1}, #2 (#3)}
\def \np#1#2#3{Nucl.~Phys.~{\bf#1}, #2 (#3)}
\def \npbps#1#2#3{Nucl.~Phys.~B (Proc.~Suppl.) {\bf#1}, #2 (#3)}
\def \pisma#1#2#3#4{Pis'ma Zh.~Eksp.~Teor.~Fiz.~{\bf#1}, #2 (#3) [JETP
Lett. {\bf#1}, #4 (#3)]}
\def \pl#1#2#3{Phys.~Lett.~{\bf#1}, #2 (#3)}
\def \plb#1#2#3{Phys.~Lett.~B {\bf#1}, #2 (#3)}
\def \pr#1#2#3{Phys.~Rev.~{\bf#1}, #2 (#3)}
\def \pra#1#2#3{Phys.~Rev.~A {\bf#1}, #2 (#3)}
\def \prd#1#2#3{Phys.~Rev.~D {\bf#1}, #2 (#3)}
\def \prl#1#2#3{Phys.~Rev.~Lett.~{\bf#1}, #2 (#3)}
\def \prp#1#2#3{Phys.~Rep.~{\bf#1}, #2 (#3)}
\def \ptp#1#2#3{Prog.~Theor.~Phys.~{\bf#1}, #2 (#3)}
\def \rmp#1#2#3{Rev.~Mod.~Phys.~{\bf#1}, #2 (#3)}
\def \rp#1{~~~~~\ldots\ldots{\rm rp~}{#1}~~~~~}
\def \si90{25th International Conference on High Energy Physics, Singapore,
Aug. 2-8, 1990}
\def \slc87{{\it Proceedings of the Salt Lake City Meeting} (Division of
Particles and Fields, American Physical Society, Salt Lake City, Utah,
1987), ed.~by C. DeTar and J. S. Ball (World Scientific, Singapore, 1987)}
\def \slac89{{\it Proceedings of the XIVth International Symposium on
Lepton and Photon Interactions,} Stanford, California, 1989, edited by M.
Riordan (World Scientific, Singapore, 1990)}
\def \smass82{{\it Proceedings of the 1982 DPF Summer Study on Elementary
Particle Physics and Future Facilities}, Snowmass, Colorado, edited by R.
Donaldson, R. Gustafson, and F. Paige (World Scientific, Singapore, 1982)}
\def \smass90{{\it Research Directions for the Decade} (Proceedings of the
1990 Summer Study on High Energy Physics, June 25 -- July 13, Snowmass,
Colorado), edited by E. L. Berger (World Scientific, Singapore, 1992)}
\def \stone{{\it B Decays}, edited by S. Stone (World Scientific,
Singapore, 1994)}
\def \tasi90{{\it Testing the Standard Model} (Proceedings of the 1990
Theoretical Advanced Study Institute in Elementary Particle Physics,
Boulder, Colorado, 3--27 June, 1990), edited by M. Cveti\v{c} and P.
Langacker (World Scientific, Singapore, 1991)}
\def \vanc{29th International Conference on High Energy Physics, Vancouver,
23--31 July 1998}
\def \yaf#1#2#3#4{Yad.~Fiz.~{\bf#1}, #2 (#3) [Sov.~J.~Nucl.~Phys.~{\bf #1},
#4 (#3)]}
\def \zhetf#1#2#3#4#5#6{Zh.~Eksp.~Teor.~Fiz.~{\bf #1}, #2 (#3) [Sov.~Phys.
-- JETP {\bf #4}, #5 (#6)]}
\def \zpc#1#2#3{Zeit.~Phys.~C {\bf#1}, #2 (#3)}


\begin{thebibliography}{99}

\bibitem{Watson} K. M. Watson, \pr{95}{228}{1954}.

\bibitem{WolfFSI} L. Wolfenstein, \prd{43}{151}{1991}.

\bibitem{DGPS} J. F. Donoghue, E. Golowich, A. A. Petrov, and J. M. Soares,
\prl{77}{2178}{1996}.

\bibitem{BSW} M. Wirbel, B. Stech, and M. Bauer, \zpc{29}{637}{1985}; M. Bauer,
B. Stech, and M. Wirbel, \zpc{34}{103}{1987}; M. Bauer and M. Wirbel,
\ibj{42}{671}{1989}.

\bibitem{BJ} J. D. Bjorken, in {\it New Developments in High-Energy Physics},
Proc.~IV International Workshop on High-Energy Physics, Orthodox Academy of
Crete, Greece, 1--10 July 1988, edited by E. G. Floratos and A. Verganelakis,
\npbps{11}{325}{1989};  D. Bortoletto and S. Stone, \prl{65}{2951}{1990}. 

\bibitem{BBNS} See, e.g., the summary by M. Beneke, preprint hep-ph/0207228,
talk at Conference on Flavor Physics and CP Violation (FPCP), Philadelphia,
Pennsylvania, 16--18 May 2002; M. Neubert, Cornell University report
CLNS-02-1794, hep-ph/0207327, invited plenary talk given at International
Workshop on Heavy Quarks and Leptons, Vietri sul Mare, Salerno, Italy, 27 May
 -- 1 Jun 2002, and at QCD 02: High-Energy Physics International Conference on
Quantum Chromodynamics, Montpellier, France, 2--9 Jul 2002.

\bibitem{KLS} Y.-Y. Keum, H.-N. Li, and A. I. Sanda, \plb{504}{6}{2001}.

\bibitem{JRcharm} J. L. Rosner, \prd{60}{114026}{1999};
for an updated analysis, see
C.~W.~Chiang, Z.~Luo and J.~L.~Rosner,
arXiv:hep-ph/0209272, to appear in Phys.\ Rev.\ D.

\bibitem{MkIII} Mark III \cn, J. Adler \ite, \plb{196}{107}{1987}.

\bibitem{Kamal} A. N. Kamal and R. C. Verma, \prd{35}{3515}{1987}, \ibj{36}
{3527(E)}{1987}.

\bibitem{Banff} J. L. Rosner, in {\it Particles and Fields 3}, Proceedings of
the Banff Summer Institute (CAP) 1988, 14--26 August 1988, edited by A. N.
Kamal and F. C. Khanna (World Scientific, Singapore, 1989), p.~395, \art.

\bibitem{SuzDB} M. Suzuki, \prd{58}{111504}{1998}.

\bibitem{JRFSI} J. L. Rosner, \prd{60}{074029}{1999}.

\bibitem{Anjos} Fermilab E691 \cn, J. C. Anjos \ite, \prd{48}{56}{1993}.

\bibitem{ARGUS} ARGUS \cn, H. Albrecht \ite, \plb{308}{435}{1993}.

\bibitem{Frab} Fermilab E687 \cn, P. Frabetti \ite, \plb{331}{217}{1994}.

\bibitem{GHLR} M. Gronau, Oscar F. Hern\'andez, D. London, and J. L. Rosner,
\prd{50}{4529}{1994}, \ibj{52}{6356, 6374}{1995}.

\bibitem{eta} M. Gronau and J. L. Rosner, \prd{53}{2516}{1996}; A. S. Dighe,
M. Gronau, and J. L. Rosner, \plb{367}{357}{1996}; \ibj{377}{325(E)}{1996}. 

\bibitem{PedlarD} CLEO \cn~results on charm production, presented by
T. Pedlar at 5th International Conference on Hyperons, Charm, and Beauty
Hadrons, Vancouver, BC, Canada, 25--29 July 2002, Proceedings edited
by C. S. Kalman \ite, to be published in Nucl.\ Phys.\ B Proc.\ Suppl.

\bibitem{Ahmed} CLEO \cn, S. Ahmed \ite, \prd{66}{031101}{2002}.

\bibitem{BelDrho} Belle \cn, A. Satpathy \ite, KEK preprint 2002-109,
hep-ex/0211022.

\bibitem{BeDK} Belle \cn, P. Krokovny \ite, preprint hep-ex/0212066.

\bibitem{Xing} Z. Z. Xing, Beijing Institute for High Energy Physics
Report No.\ BIHEP-TH-2003-1, hep-ph/0301024 (unpublished).

\bibitem{HY} H. Yamamoto, Harvard University Report No.\ HUTP-94-A006,
hep-ph/9403255 (unpublished).

\bibitem{NP} M. Neubert and A. A. Petrov, \plb{519}{50}{2001}.

\bibitem{Chau} L. L. Chau \ite, \prd{43}{2176}{1991}, \art.

\bibitem{PDG} Particle Data Group, K. Hagiwara \ite, \prd{66}{010001}{2002}.

\bibitem{BeDsK} P. Krokovny \ite, Belle \cn, \prl{89}{231804}{2002}.

\bibitem{BaDsK} B. Aubert \ite, BaBar \cn, preprint hep-ex/0211053
(unpublished).

\bibitem{DGR} A. Dighe, M. Gronau, and J. L. Rosner, \prd{57}{1783}{1988}.

\bibitem{ChauDisc} L. L. Chau and H.-Y. Cheng, \prd{39}{2788}{1989};
L. L. Chau, H.-Y. Cheng, and T. Huang, \zpc{53}{413}{1992}.

\bibitem{BFT} P. Ball, J.-M. Fr\`ere, and M. Tytgat, \plb{365}{367}{1996}.

\bibitem{WP} L. Wolfenstein, \prl{51}{1945}{1983}.

\bibitem{HJL} H. J. Lipkin, \prl{44}{710}{1980}; \ibj{46}{1307}{1981};
in {\it Proceedings of the Second International Conference on Hadron
Spectroscopy}, April 16--18, 1987, KEK, Tsukuba, Japan, edited by Y. Oyanagi,
K. Takamatsu, and T. Tsuru (National Laboratory for High Energy Physics,
Tsukuba, Japan, 1987), p.~363.

\bibitem{BGR} B. Blok, M. Gronau, and J. L. Rosner, \prl{78}{3999}{1997}.

\bibitem{HYC} Hai-Yang Cheng, preprint hep-ph/0212117 (unpublished).
\end{thebibliography}
\end{document}